\documentclass[epj]{svjour}
\usepackage{graphics}

\begin{document}
\title{Proposed search for an electric-dipole moment using laser-cooled
$^{171}$Yb atoms}
\author{Vasant Natarajan
\thanks{Electronic mail: vasant@physics.iisc.ernet.in}}
\institute{Department of Physics, Indian Institute of
Science, Bangalore 560 012, INDIA}

\date{Received: date / Revised version: date}

\abstract{We propose an experiment to search for a
permanent atomic electric-dipole moment (EDM) using
laser-cooled $^{171}$Yb atoms launched in an atomic
fountain. A uniform $B$ field sets the quantization axis,
and the Ramsey separated-oscillatory-fields method is used
to measure the Zeeman precession frequency of the atoms.
Laser beams of appropriate polarization are used for
preparation and detection in a given magnetic sublevel. The
signature of an EDM is a shift in the Ramsey resonance
correlated with application of a large $E$ field. The
precision is expected to be at least 20 times better than
current limits because the use of a cold atomic beam allows
application of $E$ field 10 times larger than in a vapor
cell, and the interaction time with the $E$ field is 200
times larger compared to a thermal beam. The leading source
of systematic error in beam experiments, the ${\bf E}
\times {\bf v}/c$ motional magnetic field, is reduced
considerably because of the near-perfect reversal of
velocity between up and down trajectories through the
$E$-field region. \PACS{
      {32.80.Pj}{Optical cooling of atoms; trapping} \and
      {32.10.Dk}{Electric and magnetic moments,
      polarizability} \and
      {11.30.Er}{Charge conjugation, parity, time reversal, and
      other discrete symmetries}
     } 
} 
\maketitle

\section{Introduction}

The existence of a permanent electric-dipole moment (EDM)
in a particle implies that both parity (P) and
time-reversal symmetry (T) are violated. Experimental
searches for an EDM are motivated by the discovery of CP
violation (and the consequent T violation) in neutral kaon
decay \cite{kaondecay}. The Standard Model (SM)
accommodates CP violation by predicting EDMs that are 8 to
9 orders of magnitude less than current experimental
limits. However, theories going beyond the SM, such as
supersymmetry, predict EDMs within experimental range
\cite{BAR93}, and are strongly constrained by measured
limits on EDMs. Thus EDM searches form an important tool in
looking for new physics beyond the SM.

Some of the most sensitive EDM searches have so far been
done on the neutron \cite{edmneutron,edmneutron2}, or on
different atoms and molecules. In atoms, an EDM can arise
due to either (i) an intrinsic EDM of the electron, (ii) an
intrinsic EDM of the neutron or proton, or (iii) a
PT-violating nucleon-nucleon or electron-nucleon
interaction \cite{KKY88}. Different atoms have different
sensitivities to these sources of EDM. In heavy
paramagnetic atoms such as Cs and Tl, the atomic EDM is
enhanced by a factor of 100--1000 times over the intrinsic
electron EDM due to relativistic effects \cite{sandars}.
Therefore experiments on such atoms put stringent limits on
the existence of an electron EDM \cite{edmsandars,MKL89}.
The best limit of $1.6 \times 10^{-27}$ e-cm comes from an
experiment using a thermal beam of $^{205}$Tl atoms
\cite{RCS02}. On the other hand, diamagnetic atoms (such as
Hg and Yb) are more sensitive to the nuclear Schiff moment
and possible PT-odd interactions. Experiments have been
done on $^{129}$Xe \cite{edmXe} and $^{199}$Hg \cite{RGJ01}
in vapor cells, with the best limit of $2.1 \times
10^{-28}$ e-cm from the $^{199}$Hg experiment. Sensitive
EDM searches have also been done using diatomic molecules
containing heavy atoms, notably TlF \cite{edmTlF} and YbF
\cite{edmYbF}. Recently, PbO has been proposed as a good
candidate for probing the electron EDM \cite{edmPbO}.

In this paper, we propose an experiment to search for an
atomic EDM using laser-cooled $^{171}$Yb launched in an
atomic fountain. The experiment is performed in the
presence of a uniform magnetic field that sets the
quantization axis. As the atoms fall under gravity, they
are first spin polarized, and then interact with an
oscillating magnetic field twice, once on the way up and
once on the way down. The interactions are used to perform
a Ramsey separated-oscillatory-fields measurement
\cite{RAM56} of the Zeeman precession frequency. In between
the two interactions, the atoms pass through a region of
large electric field, applied in a direction parallel to
the magnetic field. The signature of an EDM is a shift in
the Ramsey resonance correlated with the reversal of the
electric field.

There are three primary advantages to our scheme. The first
is that laser cooling provides a slow, dense sample of
atoms that is almost purely mono-energetic. Thus the
interaction time with the electric field is very long
(compared to a thermal beam) and almost exactly the same
for all atoms. The second advantage is that we use the
power of the Ramsey technique to measure the Zeeman
precession frequency. This ensures high precision in the
frequency measurement. The third advantage is that the
leading source of systematic error, namely the effect of
the motional magnetic field ($B_{\rm mot} = {\bf E} \times
{\bf v}/c$), is greatly reduced. This is because the
velocity of the atoms reverses between the up and down
trajectories, and the net effect is zero.

We have chosen Yb because it has been successfully laser
cooled in our laboratory \cite{RKW03} and elsewhere
\cite{KHT99,LBS00}. It is a heavy diamagnetic atom and
effects leading to EDM are comparable to that in Hg.
Furthermore, atomic calculations in Yb, which are necessary
for relating the measured EDM to fundamental PT-violating
interactions, are well developed \cite{dilip}. Precise
atomic calculations in Yb \cite{DAS97} have also been
motivated by the fact that Yb is a promising candidate for
measuring atomic parity-nonconservation effects
\cite{DEM95,KIM01}. There are two isotopes of Yb that are
suitable for an EDM measurement: $^{171}$Yb ($I=1/2$) and
$^{173}$Yb ($I=5/2$). Both isotopes have roughly the same
natural abundance (14.3\% and 16.1\%, respectively), but we
have chosen $^{171}$Yb because its simple magnetic-sublevel
structure allows for an extremely state-selective detection
scheme, as discussed later. However, it might be
interesting to measure EDM in both isotopes to address
nuclear-interaction related uncertainties when comparing
experimental results with atomic calculations.

The use of laser-cooled atoms to measure EDM has been
proposed earlier. In the early days of laser cooling,
possible measurement of the electron EDM using cold Cs
atoms in an atomic fountain was considered \cite{BVH94}.
However, it was shown that Cs has potential problems with
cold collisions that cause spin relaxation and lead to
frequency shifts \cite{GIC93}. Closed-shell fermionic atoms
with a $^1S_0$ ground state, such as $^{171}$Yb, do not
have these problems. Indeed, once these atoms are spin
polarized, the $s$-wave scattering cross section is zero,
while higher scattering cross sections go to zero at
sufficiently low temperatures. Thus the spin-coherence time
is of order 1000 s or more because effects such as
three-body recombination rates and collisional spin
relaxation are greatly reduced. There is a recent proposal
to measure EDM using laser-cooled Cs atoms trapped in
optical lattices \cite{CLV01}. In this case, there are
potential systematic effects due to AC Stark shifts in the
Zeeman sublevels caused by the trapping fields. Similar
Zeeman frequency shifts are present for proposed EDM
experiments using cold atoms in far-detuned dipole traps.
The size of these shifts have been calculated both for
paramagnetic atoms (Cs) and diamagnetic atoms (Hg)
\cite{RON99}, showing that the experimental configuration
needs to be carefully designed to control these effects.
The major advantage of a fountain experiment is that there
is no perturbation from trapping fields during the EDM
measurement. However, it should be noted that the
interaction time in a fountain experiment is limited to
less than a second due to gravity, whereas the
background-collision limited interaction time in an optical
lattice can be 10--100 times longer.

\section{Experimental details}

Atomic EDMs are measured using spin-polarized atoms in the
presence of parallel (or anti-parallel) electric (${\bf
E}$) and magnetic (${\bf B}$) fields. Since the total
angular momentum (${\bf F}$) is the only vector quantity in
the body-fixed frame, both the electric-dipole moment
(${\bf d}$) and the magnetic-dipole moment (\boldmath ${\rm
\mu}$\unboldmath) are proportional to ${\bf F}$. Therefore,
the interaction Hamiltonian in the presence of the ${\bf
E}$ and ${\bf B}$ fields is given by:
\begin{equation}
H_{\rm int} = - \left( d {\bf E} + \mu {\bf B} \right)
\cdot \frac{{\bf F}}{F} \, .
\end{equation}
The Zeeman precession frequency changes when the direction
of the ${\bf E}$ field is reversed from parallel to anti-parallel.
For $^{171}$Yb, the nuclear spin $I$ is $1/2$,
and an atom in the ${^1S}_0$ ground state has $F=1/2$. Thus
the change in the Zeeman precession frequency on $E$-field
reversal is
\begin{equation}
\Delta \omega_0 = \frac{2 d E}{\hbar} \, .
\end{equation}
Measurement of $\Delta \omega_0$ therefore constitutes a
measurement of the EDM $d$.

The above analysis shows that, in order to measure $d$
precisely, one needs to (i) measure the Zeeman precession
frequency very precisely, (ii) have a large $E$ field, and
(iii) keep the interaction time with the $E$ field as large
as possible. Atomic EDM measurements are usually performed
using thermal beams \cite{RCS02,CRD94} or in vapor cells
\cite{MKL89,RGJ01}. With thermal beams, the main limitation
is that the interaction time is only a few milliseconds
even if the $E$-field region is 100 cm long. We will see
later that the use of cold atoms increases the interaction
time by a factor of 200. In vapor-cell experiments, the
applied $E$ field is limited by the high pressure to about
10 kV/cm. The use of an atomic beam allows the $E$ field to
be at least 10 times higher.

The schematic of the proposed experiment is shown in Fig.\
1. The atoms are first laser cooled and trapped in a
magneto-optic trap (MOT). For Yb, there are two transitions
that can be used for laser cooling: the ${^1S}_0
\leftrightarrow$ ${^1P}_1$ transition at 399 nm, and the
${^1S}_0 \leftrightarrow$ ${^3P}_1$ inter-combination line
at 556 nm. Both lines are accessible with existing laser
technology, the first using a frequency-doubled Ti-sapphire
laser, and the second using a dye laser operating with
Rhodamine 110 dye. We have earlier shown that a Yb MOT can
be directly loaded from a thermal source (without the use
of a Zeeman slower) using the 399 nm line \cite{RKW03}. The
source is not isotopically enriched and contains all the
seven stable isotopes in their natural abundances. We are
able to trap each isotope individually since the isotope
shifts are about 100--1000 MHz and the individual
transitions are clearly resolved. The typical number of
trapped atoms is more than $10^8$. In Ref.\ \cite{KHT99},
Yb atoms emanating from an oven are first slowed in a
Zeeman slower using the 399 nm line, and then captured in a
MOT operating with the 556 nm line. The primary advantage
of the 556 nm line is that its natural linewidth is only
180 kHz, which results in a Doppler cooling limit of 4
$\mu$K. Therefore, for the EDM experiment it is desirable
to use a MOT with the 556 nm line. The MOT can be loaded
directly from a Zeeman slower or from another MOT operating
with the 399 nm line.

Once the atoms are loaded into the trap, the trap-magnetic
field is turned off and the atoms are allowed to
equilibriate in the optical molasses. They are then
launched upwards using the standard technique of moving
molasses: the detuning of the vertical beams is adjusted so
that the atoms are cooled in a frame moving upwards at a
velocity of 2.5 m/s. Since the energy spread in the
vertical direction still corresponds to the cooling limit
of 4 $\mu$K, the vertical velocity varies by less than 1\%.
This means that the spread in the interaction time is only
of order 1\%. By comparison, the velocity spread in the Tl
experiment is $\pm 50$\%, corresponding to the full thermal
distribution at 970 K \cite{RCS02}.

The EDM experiment is done in the presence of a static
magnetic field of 1.33 G that also sets the quantization
axis. The resulting Zeeman precession frequency in the
$F=1/2$ ground state is $\sim 1$ kHz. For comparison the Tl
experiment uses a magnetic field of 0.4 G, whereas the Hg
experiment uses a field of 0.015 G.

The freely falling atoms are first spin polarized using a
beam of right-circularly polarized ($\sigma^+$) light at
556 nm. The laser is tuned to the $F=1/2 \rightarrow
F'=1/2$ transition. Since this is a closed transition,
atoms are optically pumped into the $m_F = +1/2$ sublevel
of the ground state. The atoms then pass through an
interaction region consisting of a magnetic field rotating
at the Zeeman precession frequency. The strength of the
rotating field is adjusted such that the interaction
appears as a $\pi/2$ pulse, which puts the atoms in an
equal superposition of $m_F = +1/2$ and $m_F = -1/2$
sublevels. Since the velocity spread is $\sim 1$\%, all the
atoms experience a $\pi/2$ pulse. On the way down, the
atoms interact again with the rotating field for a second
$\pi/2$ pulse. If the oscillator is exactly on resonance
with the Zeeman precession frequency, the second $\pi/2$
pulse completes the transfer of population to the $m_F =
-1/2$ sublevel. This is a standard Ramsey
separated-oscillatory-fields method for measuring the
precession frequency \cite{RAM56}.

Population in the $m_F = -1/2$ sublevel is detected using a
probe laser on the 556 nm line. The probe laser is similar
to the optical pumping laser: it is right-circularly
polarized ($\sigma^+$) and drives the $\{F=1/2,m_F=-1/2 \}
\rightarrow \{F'=1/2,m_{F'}=+1/2 \}$ transition. As shown
in Fig.\ 2, this is an extremely state-selective detection
scheme. Any atoms in the $m_F = +1/2$ sublevel of the
ground state do not interact with the probe laser because
the only transition from this sublevel driven by the
$\sigma^+$ light is to the $\{F'=3/2,m_{F'}=+3/2 \}$
sublevel, which is almost 6 GHz away \cite{CCL79}. Thus the
laser is detuned by more than 30 000 linewidths, and the
transition probability is reduced by a factor of $10^9$.
Note that the intensity of the probe can be much greater
than the saturation intensity (0.14 mW/cm$^2$) since it is
not important that atoms absorb only once from the laser.
After the first excitation, atoms can decay back into
either sublevel and the second excitation takes place only
for those atoms that decay into the $m_F = -1/2$ sublevel.
Indeed, to maximize the signal-to-noise ratio, one would
like to have the atoms continue to interact with the probe
laser until all of them are optically pumped into the $m_F
= +1/2$ sublevel. A simple way to achieve this is to use
the state-preparation beam also as the detection beam. The
signal could be either from the absorbed photons or the
emitted fluorescence.

In between the two $\pi/2$ pulses, the atoms go through a
region of large $E$ field (where the $B$ field is also
present). Since the atoms are launched upwards with a
velocity of 2.5 m/s, the height before they turn around due
to gravity is 32 cm. Therefore the $E$-field interaction
can be about 30 cm long, corresponding to a total
interaction time of 500 ms. By comparison, the interaction
time in the Tl experiment is only 2.4 ms, even though the
$E$-field plates are 100 cm long \cite{RCS02}.

\section{Expected statistical precision}

To estimate the expected precision in the measurement, we
first note that the atoms are like oscillators whose phase
is being measured by the Ramsey technique. If the atom
starts with an initial phase $\phi_0$, then after the
Ramsey interrogation time $T$, the phase is given by
\begin{equation}
\phi = \phi_0 + \omega_0 T + \phi_n \, ,
\end{equation}
where $\phi_n$ is the additional (random) phase noise due
to spin decoherence. The presence of this noise limits the
statistical uncertainty in each measurement. If the
decoherence is a Poissonian process characterized by a time
constant $\tau$, the variance in $\phi$ increases linearly
with the measurement time as $T/\tau$. Therefore, the
frequency measurement on each atom has an uncertainty given
by $1/\sqrt{\tau T}$. If we make a simultaneous measurement
on an ensemble of $N$ atoms, then the shot-noise-limited
uncertainty in the average frequency is
\begin{equation}
\delta \omega_{SN} = \frac{1}{\sqrt{N \tau T}} \, .
\end{equation}
Even though the above equation suggests that the
measurement time $T$ should be increased indefinitely to
minimize the noise, it is clear from Eq.\ (3) that $T$
should not be much larger than $\tau$, since otherwise the
phase would be completely randomized by the noise. Rather,
the statistical error should be reduced by repeating the
measurement several times. The final error in the EDM $d$
after repeating the measurement $R$ times is:
\begin{equation}
\delta d = \frac{\hbar}{2E \sqrt{RN \tau T}} \, .
\end{equation}

To estimate the coherence time $\tau$ in the fountain, we
note that Yb is a closed-shell atom and the coherence time
is expected to be very long. For example, in the analogous
case of Hg, coherence times on the order of 500 s are
achieved in a room-temperature vapor cell by using buffer
gases (N$_2$ and CO) and having paraffin-coated cell walls
to minimize spin relaxation \cite{RGJ01}. On the other
hand, in the case of paramagnetic atoms such as Cs, the
coherence time in a vapor cell is only about 15 ms even in
the presence of N$_2$ buffer gas \cite{MKL89}. For
laser-cooled Na atoms, the coherence time measured in a
far-off resonance, red-detuned dipole trap is again only 15
ms \cite{DLA95}. However, in the same experiment, the
coherence time increased by 300 times to 4.4 s when a
blue-detuned trap of comparable depth was used. In the
blue-detuned trap, atoms are repelled by the optical
potential and spend most of their time in free fall. This
demonstrates the advantage of having the atoms away from
any trapping potential. Going by the experiments with Na, a
conservative estimate for the coherence time of Yb atoms in
a fountain is 1000 s, where we have assumed only a factor
of two increase from the value for similar atoms in a vapor
cell.

To calculate the final precision, we assume that the
applied $E$ field is 100 kV/cm, which is the typical field
used in atomic-beam experiments. The number of atoms $N$ in
the fountain can be as high as $10^9$. Using $\tau = 1000$
s and $T = 0.5$ s, the shot-noise limited uncertainty in
the frequency measurement is 225 nHz (or a relative
precision of $2.25 \times 10^{-10}$ in the measurement of
the Zeeman precession frequency). The repetition rate of
the experiment can be about 0.5 Hz. Therefore, we can reach
the precision of the Hg experiment ($2 \times 10^{-28}$
e-cm) in about 18 minutes. By contrast, the Hg experiment
takes about 300 days to achieve this precision. In 5 days,
we would achieve a precision of $1 \times 10^{-29}$ e-cm
(frequency precision of 0.484 nHz), which is 20 times
better than the best previous measurement. This analysis
also shows the advantage of using a cold atomic beam. If we
used instead a thermal beam moving at 400 m/s, the
interaction time would be 0.75 ms. Even if we assume that
the coherence time is 1000 s, the frequency uncertainty in
each measurement would be 5.8 $\mu$Hz. This is why the Tl
experiment has a frequency uncertainty as large as 25
$\mu$Hz even after 6 days of averaging.

\section{Systematic errors -- the {${\bf E} \times {\bf
v}$} effect}

The above estimate of the statistical error is meaningful
only if we can keep systematic errors below this level. The
leading source of systematic error in any beam experiment
is due to the motional magnetic field, i.e.\ the applied
electric field appears as a magnetic field in the rest
frame of the atom. If the atom is moving with a velocity
${\bf v}$, the total magnetic field in the atom's frame is
given by:
\begin{equation}
{\bf B_{\rm atom}} = {\bf B} + \frac{1}{c} {\bf E} \times
{\bf v} \, .
\end{equation}
The measured Zeeman precession frequency is proportional to
the magnitude of this field:
\begin{equation}
B_{\rm atom} = \left[ \left( {\bf B} + \frac{1}{c} {\bf E}
\times {\bf v} \right) \cdot \left( {\bf B} + \frac{1}{c}
{\bf E} \times {\bf v} \right) \right]^{1/2} \, .
\end{equation}
Using the fact that $B \gg | {\bf E} \times {\bf v} | /c$,
we expand the square root as
\begin{equation}
B_{\rm atom} \approx B +
\frac{ \left( {\bf E} \times {\bf v} \right) \cdot {\bf B}
} {cB}
+ \frac{ \left( {\bf E} \times {\bf v} \right)^2 }{2c^2B}
\, .
\end{equation}
The first term in the above equation gives the correct
Zeeman precession frequency. The third term is unimportant
because it is very small and, moreover, is even under
reversal of the $E$ field. However, the second term leads
to a systematic error since it is odd under $E$-field
reversal and mimics the EDM signal. The shift in the
precession frequency due to this term is given by:
\begin{equation}
\omega_{E \times v} = \gamma
\frac{ \left( {\bf E} \times {\bf v} \right) \cdot {\bf B}
}{cB} \, ,
\end{equation}
where $\gamma$ is the gyromagnetic ratio, equal to $4.7288
\times 10^3$ rad s$^{-1}$G$^{-1}$ for $^{171}$Yb.

It is clear from Eq.\ (9) that the systematic error cancels
if the sign of ${\bf v}$ reverses during the measurement.
In the Tl experiment, this is achieved by using two atomic
beams, one going up and the other going down, and carefully
adjusting the oven temperatures for cancellation of the
shift. The use of an atomic fountain has the velocity
reversal built into it. During the free fall, the $z$-component
of velocity at any point along the trajectory
changes sign between the upward and downward trajectories.
Since the $E$ and $B$ fields are nominally along the $x$
direction, the net effect is zero.

However, to account fully for the shift, we must consider
that the $E$ and $B$ fields may not be perfectly in the $x$
direction, and that there are small velocity components
along the $x$ and $y$ directions that do not reverse under
gravity. For this analysis, we choose our coordinate system
with the $z$ axis defined by gravity and the $x$ axis
defined by the nominal $E$ (and $B$) field direction. In
such a system, the residual components of the $E$ and $B$
fields arise mainly due to alignment errors while the
residual velocity components arise due to two reasons: (i)
finite transverse temperature in the optical molasses and
(ii) misalignment between the vertical molasses beam and
the gravity axis. To estimate the size of these effects,
let us expand Eq.\ (9):
\begin{eqnarray}
\omega_{E \times v} \approx \frac{\gamma}{cB_x}
[ & (B_xE_y - B_yE_x)v_z + (B_zE_x - B_xE_z)v_y  \nonumber
\\
&  + (B_yE_z - B_zE_y)v_x] \, .
\end{eqnarray}
The first term in the above equation is by far the dominant
term and cancels under the perfect reversal of $v_z$. The
third term is a product of two residual fields with a small
velocity component, and is completely negligible. The
second term is the only one we have to consider because it
includes the large $x$ components of the fields:
\begin{equation}
\omega_{E \times v} \approx \frac{\gamma}{c}
\left( \frac{B_z}{B_x} - \frac{E_z}{E_x} \right) E_x v_y \,
.
\end{equation}

The transverse temperature in the molasses is 4 $\mu$K,
which results in an rms velocity of $v_y = 1.4$ cm/s.
However, the mean velocity is zero, and there are equal
numbers of atoms with positive and negative velocity
components. Therefore, averaged over all atoms, the net
effect due to the transverse temperature is zero. On the
other hand, if there is a misalignment between the vertical
molasses direction and the direction of gravity, atoms
would be launched in a direction inclined to gravity and
there would be a net transverse velocity component. The
vertical molasses direction can be geometrically aligned
with gravity to better than a part in 1000. In addition,
the direction can be optimized by maximizing the number of
atoms that return to the starting point in the atomic
fountain. Assuming such alignment, we find the value of
$v_y$ is 2.5 mm/s. From Eq.\ (11), to get a systematic
shift smaller than 0.05 nHz, corresponding to an EDM of
$10^{-30}$ e-cm, the values of $B_z/B_x$ and $E_z/E_x$
should be below $2.4 \times 10^{-5}$. This is quite
reasonable since cancellation of transverse fields below $5
\times 10^{-7}$ has been reported in the Tl work
\cite{CRD94}. Furthermore, we have an experimental handle
to measure the size of this effect since $v_y$ can be
varied systematically by varying the direction of launch.
By studying its variation over $\pm 3^\circ$, one should be
able to measure the size of the term and, if necessary,
determine the point of minimum error.

One other effect of the transverse velocity is that the
atom samples a slightly different location of the field
between the up and down trajectories. Using the transverse
velocity of 2.5 mm/s, the difference in location is less
than 1 mm. Over this length scale, we can expect the field
to be uniform and safely neglect any field gradients. Again
the effect of the transverse velocity due to the finite
transverse temperature would be larger but cancels when we
average over all atoms.

We have analyzed the ${\bf E} \times {\bf v}$ effect in
detail because this is the dominant source of systematic
error in atomic-beam experiments. However, there are other
sources of error such as imperfect reversal of the $E$
field, imperfect laser-beam polarization, stray magnetic
fields from charging and leakage currents on the $E$-field
plates, stray magnetic field from the high-voltage switch,
etc. Many of these effects can be studied by the following
``experimental handles'': reversals of the $B$ field, the
phase of the oscillating field, and the polarization of the
state-preparation and detection laser. In general, the more
the number of reversals, the better the discrimination
against systematic effects. Furthermore, by varying the
vertical launch velocity and direction, one can probe
velocity-dependent systematic effects and explore different
spatial regions of the apparatus.

\section{EDM Measurement in the ${^3P}_0$ state}

There is another interesting possibility in the Yb system,
which is that the EDM can be measured in the ${^3P}_0$
metastable state. The lifetime of this state is very long
since the $0 \rightarrow 0$ transition to the ground state
is strongly forbidden. The lifetime is expected to be much
longer than the 15 s lifetime of the nearby ${^3P}_2$
state. The major advantage of the ${^3P}_0$ state is that
its mixing with states of opposite parity is 2.5 times
larger than the ground state. The nearest state of opposite
parity is the ${^3D}_1$ state, which is only 7200 cm$^{-1}$
away. This implies that PT-violating interactions leading
to an EDM will be enhanced compared to the ground state.

Experimentally, measuring EDM in the ${^3P}_0$ state of one
of the odd isotopes ($^{171}$Yb or $^{173}$Yb) is slightly
more complicated. The state would be populated using a
two-photon process (see Fig.\ 3) driving successively the
${^1S}_0 \rightarrow {^3P}_1$ transition (556 nm) and the
${^3P}_1 \rightarrow {^3D}_1$ transition ($1.54$ $\mu$m),
which results in spontaneous decay into the ${^3P}_0$ state
with a 70\% branching ratio. Detection in a magnetic
sublevel would be achieved by measuring the absorption of a
probe laser driving the ${^3P}_0 \rightarrow {^3D}_1$
transition at 1.38 $\mu$m. Diode lasers at 1.54 $\mu$m and
1.38 $\mu$m are available commercially and accessing these
transitions is not a problem. Theoretical calculations of
the enhanced EDM in the metastable state will tell us
whether the measurement is worth pursuing.

\section{Conclusions}

In summary, we have proposed a new experiment to search for
a permanent electric-dipole moment using laser-cooled
$^{171}$Yb atoms launched in an atomic fountain. Cold
diamagnetic atoms in a fountain are nearly perfect from an
experimental point of view: they are in free fall under
gravity and free from any trapping potential, they are in
an ultra-high-vacuum environment with very few collisions,
they hardly interact with each other, and they move slowly
enough that the interaction time with external fields can
be very long. We plan to use the Ramsey technique to
measure the Zeeman precession frequency in the presence of
a uniform $B$ field, which guarantees high precision in the
frequency measurement. The proposal has several advantages
compared to other measurement schemes: long interaction
times and reduction in transit-time broadening compared to
experiments using thermal beams, and use of large electric
fields compared to vapor-cell experiments. The leading
source of systematic error with atomic beams, the ${\bf E}
\times {\bf v}/c$ motional magnetic field, is greatly
reduced due to the near-perfect reversal of velocity
between up and down trajectories. Other systematic effects
that scale as the velocity should also be reduced since the
velocity is 200 times smaller compared to a thermal beam.
We estimate that a precision of $1 \times 10^{-29}$ e-cm is
achievable with 5 days integration time, which is more than
an order of magnitude better than the current limit in
$^{199}$Hg.

\section{Acknowledgments}

This work was supported by the Department of Science and
Technology, Government of India.


\begin{thebibliography}{10}

\bibitem{kaondecay}
J.~H. Christenson, J.~W. Cronin, V.~L. Fitch, and R.
Turlay, Phys. Rev. Lett.
  {\bf 13},  138–  (1964).

\bibitem{BAR93}
S.~M. Barr, Int. J. Mod. Phys. A {\bf 8},  209  (1993); W.
Bernreuther and M. Suzuki, Rev. Mod. Phys. {\bf 63}, 313
(1991).

\bibitem{edmneutron}
P.~D. Miller, W.~B. Dress, J.~K. Baird, and N.~F. Ramsey,
Phys. Rev. Lett. {\bf
  19},  381  (1967).

\bibitem{edmneutron2}
P.~G. Harris {\it et~al.}, Phys. Rev. Lett. {\bf 82},  904
(1999).

\bibitem{KKY88}
V.~M. Khatsymovsky, I.~B. Khriplovich, and A.~S.
Yelkhovsky, Ann. Phys. (Paris)
  {\bf 186},  1  (1988).

\bibitem{sandars}
P.~G.~H. Sandars, Phys. Lett. {\bf 22},  290  (1966).

\bibitem{edmsandars}
P.~G.~H. Sandars and E. Lipworth, Phys. Rev. Lett. {\bf
13},  718  (1964).

\bibitem{MKL89}
S.~A. Murthy, D. Krause, Z.~L. Li, and L.~R. Hunter, Phys.
Rev. Lett. {\bf 63},
   965  (1989).

\bibitem{RCS02}
B.~C. Regan, E.~D. Commins, C.~J. Schmidt, and D. DeMille,
Phys. Rev. Lett.
  {\bf 88},  071805  (2002).

\bibitem{edmXe}
T.~G. Vold, F.~J. Raab, B. Heckel, and E.~N. Fortson, Phys.
Rev. Lett. {\bf
  52},  2229  (1984).

\bibitem{RGJ01}
M.~V. Romalis, W.~C. Griffith, J.~P. Jacobs, and E.~N.
Fortson, Phys. Rev.
  Lett. {\bf 86},  2505  (2001).

\bibitem{edmTlF}
D. Cho, K. Sangster, and E.~A. Hinds, Phys. Rev. A {\bf
44},  904  (1991).

\bibitem{edmYbF}
J.~J. Hudson, B.~E. Sauer, M.~R. Tarbutt, and E.~A. Hinds,
Phys. Rev. Lett.
  {\bf 89},  023003  (2002).

\bibitem{edmPbO}
D. DeMille {\it et~al.}, Phys. Rev. A {\bf 61},  052507
(2000).

\bibitem{RAM56}
N.~F. Ramsey, {\em Molecular Beams} (Oxford University
Press, Oxford, 1956),
  Chap.~V.

\bibitem{RKW03}
U.~D. Rapol, A. Krishna, A. Wasan, and V. Natarajan, Eur.
Phys. J. D {\bf 29}, 409 (2004).

\bibitem{KHT99}
T. Kuwamoto, K. Honda, Y. Takahashi, and T. Yabuzaki, Phys.
Rev. A {\bf 60},
  R745  (1999).

\bibitem{LBS00}
T. Loftus, J.~R. Bochinski, R. Shivitz, and T.~W. Mossberg,
Phys. Rev. A {\bf
  61},  051401  (2000).

\bibitem{dilip}
A. Dilip {\it et~al.}, J. Phys. B {\bf 34},  3089  (2001).

\bibitem{DAS97}
B.~P. Das, Phys. Rev. A {\bf 56},  1635  (1997).

\bibitem{DEM95}
D. DeMille, Phys. Rev. Lett. {\bf 74},  4165  (1995).

\bibitem{KIM01}
D.~F. Kimball, Phys. Rev. A {\bf 63},  052113  (2001).

\bibitem{BVH94}
M. Bijlsma, B.~J. Verhaar, and D.~J. Heinzen, Phys. Rev. A
{\bf 49},  R4285
  (1994).

\bibitem{GIC93}
K. Gibble and S. Chu, Phys. Rev. Lett. {\bf 70},  1771
(1993).

\bibitem{CLV01}
C. Chin {\it et~al.}, Phys. Rev. A {\bf 63},  033401
(2001).

\bibitem{RON99}
M.~V. Romalis and E.~N. Fortson, Phys. Rev. A {\bf 59},
4547  (1999).

\bibitem{CRD94}
E.~D. Commins, S.~B. Ross, D. DeMille, and B.~C. Regan,
Phys. Rev. A {\bf 50},
  2960  (1994).

\bibitem{CCL79}
D.~L. Clark, M.~E. Cage, D.~A. Lewis, and G.~W. Greenlees,
Phys. Rev. A {\bf
  20},  239  (1979).

\bibitem{DLA95}
N. Davidson {\it et~al.}, Phys. Rev. Lett. {\bf 74},  1311
(1995).

\end{thebibliography}

\begin{figure}
\resizebox{0.95\columnwidth}{!}{\includegraphics{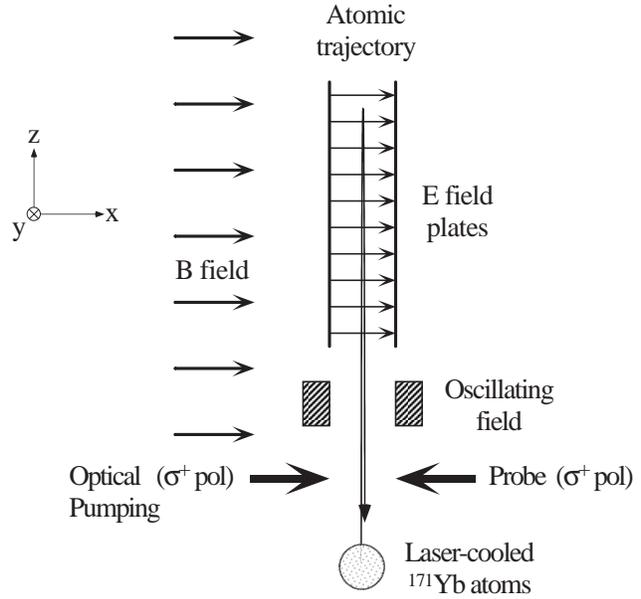}
} \caption{ Schematic of the proposed experiment.
Laser-cooled $^{171}$Yb atoms are launched upwards in an
atomic fountain. The $z$ axis is the direction of gravity.
The oscillating fields are used to measure the Zeeman
precession frequency in the $B$ field. During the free
fall, the atoms pass through a region of large $E$ field
that shifts the precession frequency by an amount
proportional to the EDM.} \label{f1}
\end{figure}

\begin{figure}
\resizebox{0.95\columnwidth}{!}{\includegraphics{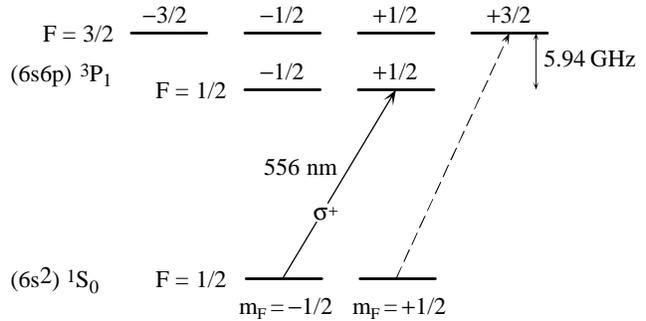}
} \caption{ State detection scheme. Atoms in the $m_F =
-1/2$ sublevel are selectively detected using
right-circularly polarized ($\sigma^+$) light at 556 nm.
The light is tuned to the $F=1/2 \rightarrow F'=1/2$
hyperfine transition. Atoms in the the $m_F = +1/2$
sublevel are not detected because the transition driven by
$\sigma^+$ light (shown by the dashed line) is to the
$F'=3/2,m_{F'}=+3/2$ sublevel, which is 5.94 GHz away.}
\label{f2}
\end{figure}

\begin{figure}
\resizebox{0.95\columnwidth}{!}{\includegraphics{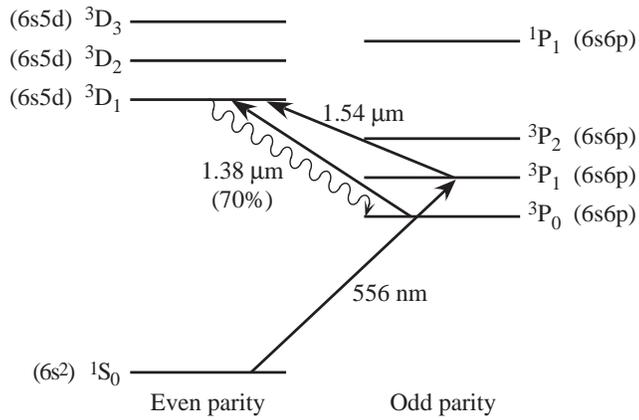}}
\caption{Low-lying energy levels of Yb (not to scale)
showing the two-step process for populating the metastable
${^3P}_0$ state. The branching ratio for decay from the
${^3D}_1$ state into the ${^3P}_0$ state is 70\%. The
experiment would use an odd isotope with additional
hyperfine structure (not shown). State detection is by
excitation back to the ${^3D}_1$ state.} \label{f3}
\end{figure}

\end{document}